\documentclass[twocolumn,aps,pre,superscriptaddress,nofootinbib]{revtex4-2}

\usepackage{pifont}

\usepackage{color, colortbl}
\definecolor{Gray}{gray}{0.2}

\usepackage{graphicx}
\usepackage[dvipsnames,svgnames]{xcolor}
\usepackage{bm,bbm}
\usepackage{braket}
\usepackage{amsthm,amsmath,amssymb}
\usepackage{enumerate}
\usepackage{booktabs,multirow}
\usepackage{mathtools,bbding}
\usepackage[percent]{overpic}
\usepackage{microtype}
\usepackage{array,adjustbox,afterpage}
\usepackage[T1]{fontenc}
\usepackage[utf8]{inputenc}
\usepackage{tikz}
\usetikzlibrary{calc}
\usepackage[english]{babel}
\usepackage{newtxtext,newtxmath}
\usepackage{dsfont}

\usepackage{lipsum}

\usepackage[linesnumbered,ruled]{algorithm2e}  %delete ruled and write boxed for box

\makeatletter
\renewcommand{\fnum@figure}{FIG. \thefigure} % Figure --> FIG.
\makeatother

\usepackage[bookmarks=false, colorlinks=true, linkcolor=blue, citecolor=purple, urlcolor=purple]{hyperref}
\usepackage{cleveref}
\usepackage[normalem]{ulem}

\usepackage{lipsum}
\usepackage{makecell}

\Crefname{subfigures}{figure}{figures}
\Crefname{subfigures}{Figure}{Figures}

\newcommand{\xdownarrow}[1]{{\left\downarrow\vbox to #1{}\right.\kern-\nulldelimiterspace}} % for long downarrow ..

\hypersetup{
    breaklinks=true,   % splits links across lines
    colorlinks=true,   % displays links as colored text instead of blocks
    pdfusetitle=true,  % \title and \author values into pdf metadata etc.
}

\begin{document}

\title{Biorthogonal resource theory of genuine quantum superposition}
\author{Onur Pusuluk}
\email{onur.pusuluk@gmail.com}
\affiliation{Faculty of Engineering and Natural Sciences, Kadir Has University, 34083, Fatih, Istanbul, T\"{u}rkiye}

\begin{abstract}
The phenomenon of quantum superposition manifests in two distinct ways: it either spreads out across non-orthogonal basis states or remains concealed within their overlaps. Despite its profound implications, the resource theory of superposition often neglects the quantum superposition residing within these overlaps. However, this component is intricately linked to a form of state indistinguishability and can give rise to quantum correlations. In this paper, we introduce a pseudo-Hermitian representation of the density operator, wherein its diagonal elements correspond to biorthogonal extensions of Kirkwood-Dirac quasi-probabilities. This representation provides a unified framework for the inter-basis quantum superposition and basis state indistinguishability, giving rise to what we term as \textit{genuine} quantum superposition. Moreover, we propose appropriate generalizations of current superposition measures to quantify genuine quantum superposition that serves as the fundamental notion of nonclassicality from which both quantum coherence and correlations emerge. Finally, we explore potential applications of our theoretical framework, particularly in the quantification of electron delocalization in chemical bonding and aromaticity.
\end{abstract}

\maketitle

\section{Introduction}\label{Sec::Intro}

Quantum coherence is regarded as a basis-dependent notion of nonclassicality in view of its definition as the quantum superposition with respect to a fixed orthonormal basis. On the other hand, quantum correlations, encompassing quantum entanglement~\cite{2009HorodeckiEnt} and more broadly quantum discord~\cite{Vedral-2001, Zurek-2002, 2010_PRL_VV_CorrWithRelEnt}, remain invariant under local unitary transformations. However, their essence is solely the quantum coherence shared in composite systems. In a sense, these quantities appears as basis-independent manifestations of quantum coherence~\cite{2015_PRA_CoherenceAndDiscord, 2016_PRA_CorrelatedCoherenceAndDiscordAndEnt, 2017_PRA_DiscordlikeCorrOfCoh, 2018_PhysicsReports_CorrelatedCoherenceAndGeoDiscord, 2019_PRA_PartialCoherence}. Bell states, which exhibit maximum bipartite entanglement, serve as an illustrative example. Within these states, quantum coherence is entirely distributed over the joint system without being localized in subsystems, irrespective of local basis choices.

\begin{figure}[b] \centering
        \includegraphics[width=.305\textwidth]{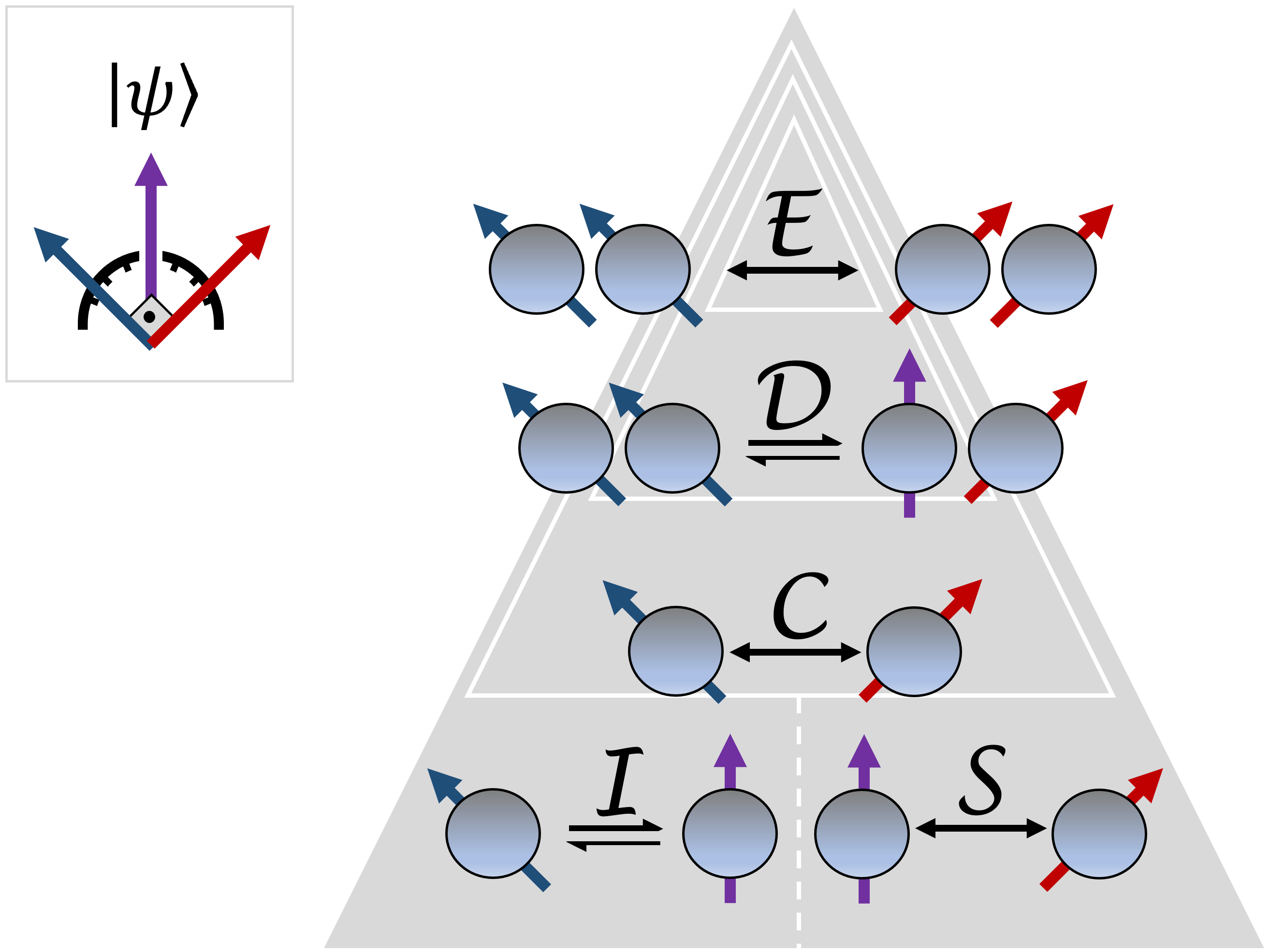}
        \caption{Hierarchy of nonclassicality between quantum entanglement \(E\), discord \(D\), coherence \(C\), superposition \(S\), and state indistinguishability \(I\). The balls depict physical systems, while the diagonal and vertical arrows represent pure physical states. The angle between the arrows quantifies the overlap between the represented states such that perpendicular arrows correspond to orthogonal states. \(\rightleftharpoons\) and \(\leftrightarrow\) stand for a probabilistic mixture (or incoherent superposition) and coherent superposition of the left and right states, respectively.}
        \label{Fig::Piramit}
\end{figure}

Interesting questions arise in any attempt to extend this unified framework to include quantum superposition as shown in Fig.~\ref{Fig::Piramit}. Does quantum superposition conceptually contain quantum coherence and quantum correlations as particular subsets? If so, can we argue that an entangled or discordant state must possess not only non-zero coherence but also non-zero superposition in any basis achievable through local unitary transformations? Otherwise, should we seek other notions of nonclassicality that can give rise to quantum coherence and correlations apart from quantum superposition? Here, we will approach these questions using quantum resource theories~\cite{2016_QRT, 2019_QRT, TJP2023}.

Each resource theory categorizes all conceivable quantum states into two groups: \textit{resource} and \textit{free}, based on their behavior under a defined set of restricted quantum operations permitted for state manipulation. For example, in the resource theory of entanglement~\cite{2009HorodeckiEnt}, free states are separable, and it remains impossible to generate an entangled state from these states using solely local operations and classical communication. Similarly, in the resource theory of quantum superposition, incoherent mixtures such as those depicted below:
\begin{equation} \label{Eq::SFreeState}
\hat{\rho}_f = \sum_i p_i |c_i\rangle \langle c_i | ,
\end{equation}
are regarded as superposition-free states~\cite{2017_PRL_Plenio, 2021_PRA_GT, 2022_PRA_GT}, where \(\{|c_i\rangle\}\) represents a set of normalized and linearly independent states that are not necessarily orthogonal, and \(\{p_i\}\) form a probability distribution. Consequently, a system existing in state \(\hat{\rho}_f\) cannot serve as a resource under quantum operations incapable of creating or amplifying quantum superposition in the basis of \(\{|c_i\rangle\}\).

\begin{figure}[t] \centering
        \includegraphics[width=.35\textwidth]{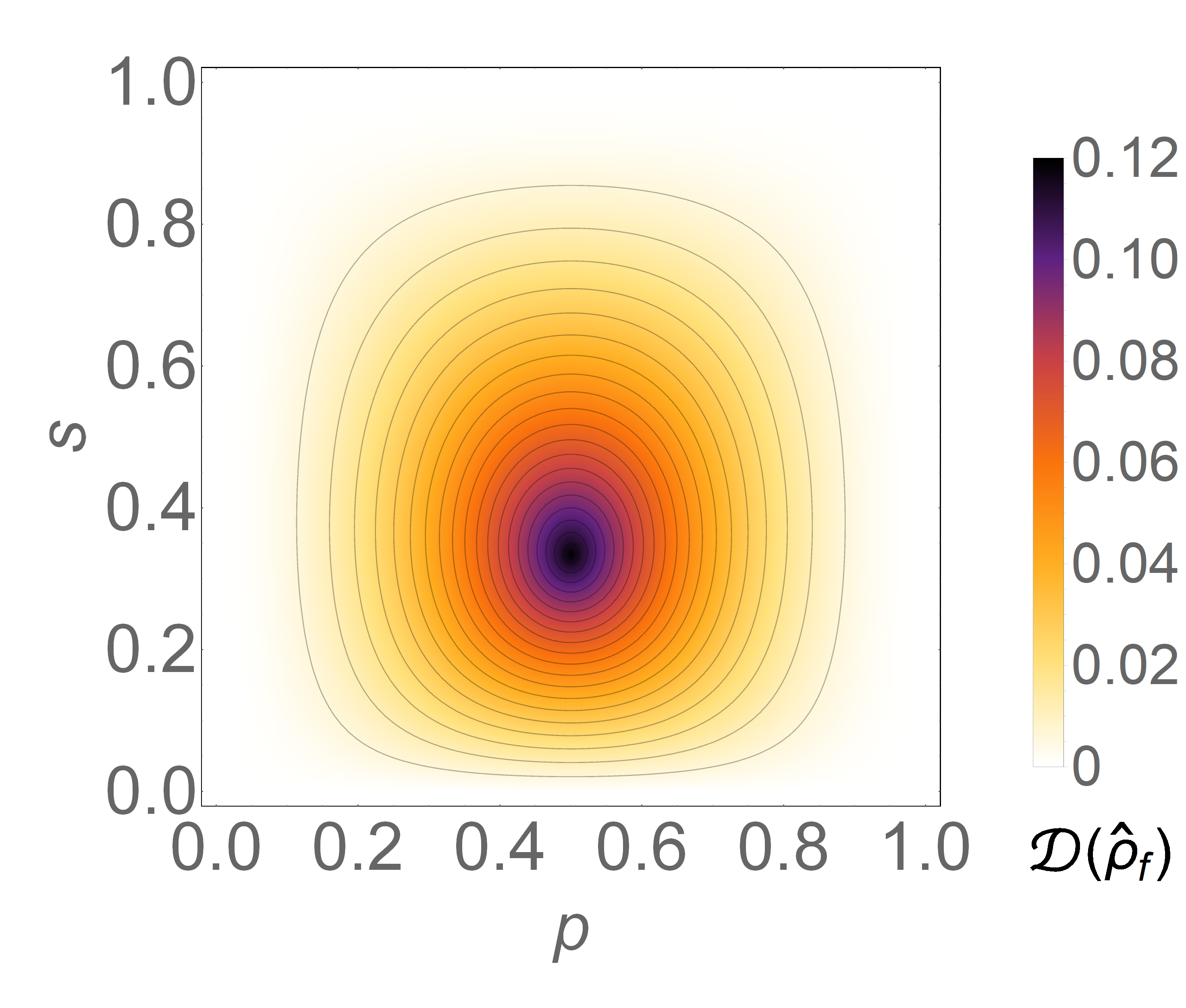}
        \caption{Geometric discord~\cite{2010_PRL_GeoDiscord} of state~(\ref{Eq::QCState}). Right and left discords are equal to each other, while the entanglement is always zero.}
        \label{Fig::GeoDiscord}
\end{figure}

For the sake of simplicity, let us consider a four-level system existing in a mixture that contains only two states \(|c_1\rangle\) and \(|c_2\rangle\) with \(\langle c_1|c_2\rangle = s \in \mathds{R}\). We will focus on a particular bipartition of this system into two two-level subsystems \(A\) and \(B\) such that \(|c_i\rangle \mapsto |a_i\rangle_A \otimes |b_i\rangle_B\) with \(\langle a_1|a_2\rangle = \langle b_1|b_2\rangle = \sqrt{s}\). Then,
\(\hat{\rho}_f\) reads
\begin{equation} \label{Eq::QCState}
p \, |a_1\rangle \langle a_1 | \otimes |b_1\rangle \langle b_1 | + (1 - p) \, |a_2\rangle \langle a_2 | \otimes |b_2\rangle \langle b_2 | \, ,
\end{equation}
which is an example of quantum-quantum states. Despite appearing as a superposition-free state in the basis of \(\{|c_1\rangle, |c_2\rangle\}\), this state possesses nonclassical correlations in the form of quantum discord (please see Fig.~\ref{Fig::GeoDiscord}).

A complete resource theory of quantum discord has yet to emerge in the existing literature. Nonetheless, it is established that incoherent operations cannot create discordant states such as~(\ref{Eq::QCState}) without consuming local quantum coherences initially present in the subsystems~\cite{2016_PRL_Benjamin}. That is to say, the so-called superposition-free state is not coherence-free in any orthogonal basis accessible through local unitary transformations. Therefore, the current resource-theoretical definition of quantum superposition appears inadequate, in a notional sense, to cover quantum coherence and quantum correlations as a subset.

As a matter of fact, the correlations shared in state~(\ref{Eq::QCState}) arise from the local indistinguishability in subsystems. With a complete measurement on \(A\) (\(B\)), the states  \(|a_1\rangle\) and \(|a_2\rangle\) (\(|b_1\rangle\) and \(|b_2\rangle\)) cannot be perfectly distinguished from each other. This makes the global system sensitive to the local dynamics. Should we take into account the basis state indistinguishability as an independent notion of nonclassicality from which the quantum coherence and correlations can arise as in Fig.~\ref{Fig::Piramit}? Or, should we extend the conventional definition of quantum superposition to include this kind of quantum indistinguishability as a special case? This is what we would like to start discussing in this paper.

To set the ground for this discussion, we propose utilizing a pseudo-Hermitian matrix representation of a quantum state to investigate its properties concerning a nonorthogonal basis. This approach entails expressing the state within the biorthogonal extension of the given basis, a well-established mathematical technique in the literature (see, for example, Refs.~\cite{1960_NuclearP, 1967_NuclearP, 1971_AOs, 1973_AOs, 1975_AOs, 1986_AOs, 2002_trio_AM, 2003_AM_jmp_03, 2010_ps_AM, 2010_ijgmmp_AM, 2014_boQM, 2019_nhQM}). Although independently rediscovered to represent free operations in the resource theory of superposition~\cite{2017_PRL_Plenio, 2021_PRA_GT, 2022_PRA_GT}, this method has not been extended to encompass quantum states. Here, we will demonstrate that the diagonal elements of the pseudo-Hermitian representation of quantum states are biorthogonal Kirkwood-Dirac quasi-probabilities~\cite{1933_PhysRev_K, 1945_RevModPhys_Dirac, 1957_PhysRev_Barut, 1961_Terletsky-Margenau-Hill, Review_2024_arXiv_Nicole}. We will introduce the term \textit{genuine quantum superposition} by associating the remaining matrix elements with both quantum superposition and basis state indistinguishability. Subsequently, we will generalize existing superposition measures for genuine quantum superposition. The measures introduced herein hold promise for quantifying non-classicality in chemical bonding phenomena within the realm of quantum chemistry, as evidenced by recent research where we successfully calculated electron delocalization in aromatic molecules~\cite{2023_Aromatiq}.

\section{Definition and Motivation} \label{Sec::Trace}

A generic density operator that lives in a \(d\)-dimensional Hilbert space can be expressed in the form of
\begin{equation} \label{Eq::State::RTS}
\hat{\rho} = \sum_{j,k=1}^d \rho_{jk} |c_j\rangle \langle c_k | ,
\end{equation}
where \(\{|c_j\rangle\}\) with \(\langle c_j|c_k\rangle = G_{jk}\) constitute a nonorthogonal but complete and minimal basis.

\textbf{Theorem 1.} In the case of nonorthogonal basis states, the trace of an operator is calculated by
\begin{equation} \label{Eq::Trace::Closed}
\mathrm{tr}[\hat{\rho}] = \sum_{i=1}^d \langle c_i^\perp | \hat{\rho} | c_i \rangle ,
\end{equation}
where \(\{|c_i^\perp\rangle\}\) with \(\langle c_i^\perp |c_j \rangle = \delta_{i,j}\) is called the \textit{dual} of the basis \(\{|c_i\rangle\}\): 
\begin{equation} \label{Eq::Transform2BiO}
|c_j^\perp \rangle = \sum_i (G^{-1})_{ij} |c_i\rangle .
\end{equation}

The two dual bases are jointly called \textit{biorthogonal}. When all the overlaps go to zero, the dual basis becomes identical to \(| c_i\rangle\) and consequently, Eq.~(\ref{Eq::Trace::Closed}) reduces to the conventional definition of trace operation in orthogonal basis.
\newline

\textbf{Proof.} This trace definition can be simply verified considering the pure superposition states
\begin{equation} \label{Eq::PureState}
|\psi\rangle = \sum_j \psi_j |c_j\rangle \, ,
\end{equation}
for which  \(\mathrm{tr}[|\psi\rangle \langle \psi |] = \langle \psi |\psi\rangle\) becomes \(\sum_{jk} \psi_j \psi_k^* \langle c_k | c_j \rangle = \sum_i \langle c_i^\perp |\psi\rangle \langle \psi | c_i \rangle\). Extending the same verification to mixed states is equally straightforward 
\begin{equation} \label{Eq::Trace::Proof}
\begin{aligned}
\mathrm{tr}[\hat{\rho}] &= \sum_{jk} \rho_{jk} \mathrm{tr}[\mathds{1} |c_j\rangle \langle c_k|] = \sum_{ijk} \rho_{jk} \langle c_i^\perp |c_j\rangle \langle c_k | c_i \rangle \\
&= \sum_{i} \langle c_i^\perp | \hat{\rho} | c_i \rangle ,
\end{aligned}
\end{equation}
as the identity operator can be expressed as
\begin{equation} \label{Eq::Identity}
\mathds{1} = \sum_i |c_i\rangle \langle c^\perp_i| \, . 
\end{equation}

\textbf{Corollary 2}. Consider a density operator written in a nonorthonormal basis \(\{|c_i\rangle\}\) as in Eq.~(\ref{Eq::State::RTS}). The overall nonclassicality of this state can be quantified by the \(l_1\) norm of the operator, defined by
\begin{equation} \label{Eq::l1norm::Closed}
l_1[\hat{\rho}] = \sum_{j\neq j} | \langle c_j^\perp | \hat{\rho} | c_k \rangle | .
\end{equation}

To ensure a density operator accurately represents a physical state, it must have a unit trace, which expresses the fact that probabilities sum to one. Consequently, Eq.~(\ref{Eq::Trace::Closed}) implies that the sum of the elements \(\langle c_i^\perp | \hat{\rho} | c_i \rangle\) accounts for the total probability. When \(j\neq k\), \(\langle c_j^\perp | \hat{\rho} | c_k \rangle\) holds no contribution to the total probability. Thus, these non-probabilistic components within the density operator, often complex numbers, signify the nonclassical characteristics of the state under examination.

\section{Overview of the Current Framework} \label{Sec::RTS}

\subsection{Mathematical Characterization} \label{Sec::RTS::Mat}

In the existing literature~\cite{2017_PRL_Plenio, 2021_PRA_GT, 2022_PRA_GT}, the quantification of quantum superposition within the density operator \(\hat{\rho}\) in the basis of \(\{|c_i\rangle\}\) typically involves examining the off-diagonal elements of the matrix defined as
\begin{equation}\label{Eq::RhoNO}
  \rho_{N\!O} \equiv \{\rho_{jk}\} = \{\langle c_j^\perp | \hat{\rho} | c_k^\perp \rangle\}.
\end{equation}
One common approach is to use the \(l_1\) norm of this matrix:
\begin{equation} \label{Eq::l1norm::RTS}
l_1[\rho_{N\!O}] = \sum_{j\neq k} |\rho_{jk}|,
\end{equation}
which becomes zero for the superposition-free states given in Eq.~\eqref{Eq::SFreeState}. Hereafter, we will refer to the matrix \(\rho_{N\!O}\) as the nonorthogonal matrix representation of \(\hat{\rho}\). However, the diagonal elements of this matrix are not sufficient to calculate the trace of the operator. This becomes apparent when \eqref{Eq::Trace::Closed} is rewritten as
\begin{equation} \label{Eq::Trace::RTS}
\mathrm{tr}[\hat{\rho}]  = \mathrm{tr}[\rho_{N\!O} \, G]  .
\end{equation}

Thus, additional information encoded in the so-called \textit{Gram} or \textit{overlap matrix} \(G\) is required to determine the total probability in the state \(\hat{\rho}\). Moreover, \(\{|c_i\rangle\}\) does not form a proper basis for investigating operations on \(\hat{\rho}\). Let \(\hat{A}\) denote another operator with matrix elements \(A_{ij}\) in the basis \(\{|c_i\rangle\}\), i.e., \(\hat{A} = \sum_{ij} A_{ij} |c_i\rangle \langle c_j|\), where \(A_{N\!O} \equiv \{A_{ij}\}\). Consequently, the action of this operator on the state, \(\hat{A} \, \hat{\rho} = \sum_{ijkl} A_{ij} \, G_{jk} \, \rho_{kl} |c_i\rangle \langle c_l|\) cannot be represented by the matrix product \(A_{N\!O} \, \rho_{N\!O}\) whose elements are \(\sum_{j} A_{ij} \, \rho_{jk}\).

Due to these practical challenges, the superposition-free operators proposed in Refs.~\cite{2017_PRL_Plenio, 2021_PRA_GT, 2022_PRA_GT} are expressed as follows:
\begin{equation}\label{Eq::OpsRTS}
  \Phi (\hat{\rho}) = \sum_n \hat{K}_n \, \hat{\rho} \, \hat{K}_n^\dagger ,
\end{equation}
where the Kraus operators \(\hat{K}_n\) are constructed in the biorthogonal basis \(\{|c_i\rangle, |c^\perp_i\rangle\}\) :
\begin{equation}\label{Eq::KrausOpsRTS}
  \hat{K}_n = \sum_{i} c_{i,n} |c_{f_n(i)} \rangle \langle c_i^\perp| ,
\end{equation} 
with \(c_{i,n}\) being complex numbers and \(f_n(i)\) representing index functions.

\subsection{Physical Interpretation} \label{Sec::RTS::Phys}

In addition to the practical challenges in its mathematical characterization, the nonorthogonal matrix representation \(\rho_{N\!O}\) also presents operational difficulties. At first glance, Eq.~(\ref{Eq::l1norm::RTS}) seems to be an extension of the \(l_1\) norm of quantum coherence~\cite{Plenio-2014} for the case when the orthogonality of basis states breaks down. When \(G_{ij}\) vanishes for all \(i \neq j\), \(\rho_{N\!O}\) corresponds to a valid density matrix that is Hermitian, positive-semidefinite, and has unit trace. Its diagonal and off-diagonal elements capture the classical and nonclassical aspects of the information encoded in the density operator; they correspond to probabilities and quantum coherences in this limit. Extending the decomposition of \(\rho_{N\!O}\) beyond this limit necessitates the consideration of a specific generalized measurement, known as a positive operator-valued measure (POVM), performed on state \(\hat{\rho}\) and described by \(d+1\) elements such that
\begin{equation}\label{Eq::POVM}
\hat{F}_i = \begin{cases}
|c^\perp_i\rangle \langle c^\perp_i| & \text{for } i \leq d \\
\mathds{1} - \sum_{j=1}^d |c^\perp_j\rangle \langle c^\perp_j| & \text{for } i = d + 1
\end{cases} \, .
\end{equation}

Then, \(\rho_{ii}\) becomes equal to \(\mathrm{tr}[\hat{\rho} \, \hat{F}_i]\), which corresponds to the probability of obtaining measurement outcome \(i\). The outcome \(i\leq d\) certifies the projection of the state \(\hat{\rho}\) onto \(|c_i\rangle\). Conversely, if the outcome is \(d + 1\), the post-measurement state remains a superposition of the basis states. The probability of this outcome is given by \(1 - \mathrm{tr}[\rho_{N\!O}]\). Hence, the POVM operation, as defined in Eq.~(\ref{Eq::POVM}), decomposes state \(\hat{\rho}\) into the basis states \(|c_i\rangle\) with a total probability of \(\mathrm{tr}[\rho_{N\!O}]\). In other words, it converts \(n\) independent copies of state \(\hat{\rho}\) into \(n \, \mathrm{tr}[\rho_{N\!O}]\) basis states and \(n \, (1 - \mathrm{tr}[\rho_{N\!O}])\) superposition states.

Unless the density operator is superposition-free, \(\mathrm{tr}[\rho_{N\!O}] \neq \mathrm{tr}[\hat{\rho}] = 1\). Thus, the so-called superposition-free states in Eq.~(\ref{Eq::SFreeState}) are the only ones perfectly decomposable into non-orthogonal states \(|c_i\rangle\) by using the POVM elements \(\{\hat{F}_i\}_{i=1}^{d+1}\). Moreover, the same quantum operation projects the pure states given in Eq.~(\ref{Eq::PureState}) into \(\sum_{j,k} f(\{\psi_j\}, \{(G^{-1})_{kj}\}) |c_k\rangle\) with a probability of \(1 - \mathrm{tr}[\rho_{N\!O}]\). In this case, the off-diagonals of \(\rho_{N\!O}\) contains no information about the \textit{residual} superposition associated with the coefficients \(f(\{\psi_j\}, \{(G^{-1})_{kj}\})\). Essentially, the \(l_1\) measure given in Eq.~(\ref{Eq::l1norm::RTS}) excludes this residual superposition, which originates from the indistinguishability of the basis states. We will catogorize the superpositions that can and cannot be quantified by \(l_1[\rho_{N\!O}]\) as \textit{inter-basis} and \textit{intra-basis} superpositions in what follows.

When we relate the elements of \(\rho_{N\!O}\) to not just one, but two distinct measurements, we encounter a similar challenge. The Kirkwood-Dirac (KD) quasi-probabilities~\cite{1933_PhysRev_K, 1945_RevModPhys_Dirac, 1957_PhysRev_Barut, 1961_Terletsky-Margenau-Hill, Review_2024_arXiv_Nicole} for the elements \(\{\hat{F}_i\}_{i=1}^{d+1}\) can be defined as follows:
\begin{equation}\label{Eq::KD4RTS}
  Q_{jk}^{N\!O} = \mathrm{tr}[\hat{F}_k \, \hat{F}_j \, \hat{\rho}] .
\end{equation}

Even if both measurements yield results less than \(d+1\), these quasi-probabilities are equal to \(\langle c^\perp_k | c^\perp_j \rangle \, \rho_{jk} = (G^{-1})_{kj} \, \rho_{jk}\), indicating that they cannot be derived solely from the elements of \(\rho_{N\!O}\). Furthermore, \(\sum_{j=1}^{d+1} Q_{jk}^{N\!O} = \rho_{k^\prime k^\prime} + 1 - \mathrm{tr}[\rho_{N\!O}]\), where \(k^\prime\) means \(k \leq d\). The marginals of KD distribution manifest as genuine probability distributions, and the diagonal elements of \(\rho_{N\!O}\) cannot be associated with these probabilities without consideration of \(1 - \mathrm{tr}[\rho_{N\!O}]\).

In the light of the discussion above, it seems that the connection between the (off-)diagonals of \(\rho_{N\!O}\) and all the (non-)probabilistic information encoded by \(\hat{\rho}\) requires further clarification. To this aim, we shall explore alternative matrix representations of \(\hat{\rho}\) whose diagonal elements are also related to probabilistic interpretations of measurement outcomes.

\section{Framework of Biorthogonality} \label{Sec::BORTS}

The utilization of the biorthogonal basis \(\{|c_i\rangle, |c^\perp_i\rangle\}\) has been previously documented in representing observables within the valence bond theory of chemical bonding~\cite{1971_AOs, 1973_AOs, 1975_AOs}, offering significant practical simplifications. Additionally, as demonstrated in the seminal work of Ref.~\cite{1986_AOs}, this basis reduces the computational complexity involved in calculating transition density matrices. Furthermore, investigating non-Hermitian quantum systems through the framework of biorthogonality presents a promising avenue across a spectrum of disciplines, including linked-cluster expansions in nuclear physics~\cite{1960_NuclearP, 1967_NuclearP}, \(\mathcal{PT}\)-symmetry and pseudo-Hermiticity~\cite{2002_trio_AM, 2003_AM_jmp_03, 2010_ps_AM, 2010_ijgmmp_AM}, and no-go theorems in quantum information theory~\cite{2019_nhQM}.

Here, we propose a pseudo-Hermitian representation of the Hermitian density operator \(\hat{\rho}\) by expressing it in the biorthogonal basis as below
\begin{equation} \label{Eq::State::BiO}
\begin{aligned}
\hat{\rho} &= \mathds{1} \, \hat{\rho} \, \mathds{1} = \Big(\sum_{j=1}^d |c_j\rangle \langle c_j^\perp |\Big) \hat{\rho} \Big(\sum_{k=1}^d |c_k\rangle \langle c_k^\perp |\Big) \\
&= \sum_{j,k=1}^d \bar{\rho}_{jk} |c_j\rangle \langle c_k^\perp | ,
\end{aligned}
\end{equation}
where the coefficients \(\bar{\rho}_{jk}\) constitute a non-Hermitian but trace-one matrix whose eigenvalues are real:
\begin{equation}\label{Eq::RhoBO}
  \rho_{BO} \equiv \{\bar{\rho}_{jk}\} = \{\langle c^\perp_j | \hat{\rho} | c_k \rangle\}.
\end{equation}

Let us call this pseudo-Hermitian matrix the \textit{biorthogonal matrix representation} of \(\hat{\rho}\). The two matrix representations in question are interconnected by:
\begin{equation}\label{Eq::NO2BO}
  \rho_{BO} = \rho_{N\!O} \, G ,
\end{equation} 
due to the following relationship between \eqref{Eq::State::RTS} and \eqref{Eq::State::BiO}:
\begin{equation}
\begin{aligned}
  \hat{\rho} &= \sum_{i,j=1}^d \rho_{ij} |c_i\rangle \langle c_j | \Big(\sum_{k=1}^d |c_k\rangle \langle c_k^\perp |\Big) \\
  &= \sum_{i,k=1}^d \Big(\sum_{j=1}^d \rho_{ij} \, G_{jk} \Big) |c_i\rangle \langle c_k^\perp | = \sum_{i,k=1}^d \bar{\rho}_{ik} |c_i\rangle \langle c_k^\perp | .
\end{aligned}
\end{equation}

Eqs.~\eqref{Eq::Trace::RTS}~and~\eqref{Eq::NO2BO} then imply that \(\mathrm{tr}[\rho_{BO}] = \mathrm{tr}[\hat{\rho}] = 1\).  Also, the diagonal elements of \(\rho_{BO}\) quantify the relative weights of the basis states \(|c_i\rangle\) in the density operator \(\hat{\rho}\).

\subsection{Pure Superposition States}

For pure superposition states~(\ref{Eq::PureState}), the diagonal elements of \(\rho_{BO}\) are identical to Chirgwin-Coulson weights~\cite{1950_Coulson}
\begin{equation} \label{Eq::CCweight}
\bar{\rho}_{ii} = \sum_{j} \psi_i \psi_j^* G_{ji} \equiv w_i \, ,
\end{equation}
which form one of the most conventional schemes for assigning nonorthogonal weights in the valance bond theory of chemical bonding~\cite{1998_VBweights}.

Moreover, \(w_i\) correspond to the elements of the vectors created to study majorization relations in order to investigate pure state transformations in the resource theory of superposition~\cite{2021_PRA_GT}, i.e., superposition-free operations cannot transform \(\sum_i \psi_i |c_i\rangle\) into \(\sum_i \psi^\prime_i |c_i\rangle\) if \((w_1, \cdot\cdot\cdot, w_d)^T \nprec (w^\prime_1, \cdot\cdot\cdot, w^\prime_d)^T\).

As observed, the diagonal elements of the biorthogonal matrix representation of pure states have been associated with probabilities in various fields ranging from quantum chemistry to quantum information. Now, we shall extend this relationship to mixed states.

\subsection{Mixed Superposition States}

Unlike \(\{\rho_{ii}\}\), \(\{\bar{\rho}_{ii}\}\) are normalized to one not only in the limit of orthogonality but also in general. Also, they can be written as \(\mathrm{tr}[\hat{\rho} \, \hat{\Pi}_i]\), where \(\hat{\Pi}_i = |c_i\rangle \langle c^\perp_i|\) satisfy the following relations
\begin{align} \label{Eq::NOProjectors}
\sum_{i=1}^d \hat{\Pi}_i = \mathds{1} \, , \quad
\hat{\Pi}_i \, \hat{\Pi}_j = \delta_{ij} \, \hat{\Pi}_i \, , \quad
\hat{\Pi}_i \, |c_j\rangle = \delta_{ij} \, |c_j\rangle \, .
\end{align}

Based on these relations, it can be conjectured that the non-Hermitian operators \(\{\hat{\Pi}_i\}\) likely serve as projection operators within biorthogonal systems~\cite{2014_boQM}, effectively defining a \textit{biorthogonally generalized} projection-valued measurement operation. This operation projects the density operator \(\hat{\rho}\) into the basis state \(|c_i\rangle\) with a probability of \(\bar{\rho}_{ii}\). Furthermore, there isn't any residual superposition in this alternative decomposition process. Hence, all the (non-)probabilistic information encoded by \(\hat{\rho}\) are presumably reflected in \(\rho_{BO}\) by its (off-)diagonal elements.

We can also successfully associate the elements of \(\rho_{BO}\) with two distinct measurements. Specifically, for biorthogonal projections operators, the KD quasi-probabilities~\cite{1933_PhysRev_K, 1945_RevModPhys_Dirac, 1957_PhysRev_Barut, 1961_Terletsky-Margenau-Hill, Review_2024_arXiv_Nicole} take the form:
\begin{equation}\label{Eq::KD4BORTS}
\begin{aligned}
  Q_{jk}^{BO} &= \mathrm{tr}[\hat{\Pi}_k \, \hat{\Pi}_j \, \hat{\rho}] \\
              &= \langle c_j^\perp | \hat{\rho} | c_k \rangle \, \langle c_k^\perp | c_j \rangle = \delta_{k,j} \, \bar{\rho}_{jk} .
\end{aligned}
\end{equation}

In essence, the biorthogonal KD quasi-probabilities are either identical to \(\bar{\rho}_{jj}\) or are null. Conversely, the diagonal elements of \(\rho_{BO}\) correspond to biorthogonal KD quasi-probabilities distinct from zero. Furthermore, its off-diagonal elements are linked to vanishing quasi-probabilities. Also, it is evident that the marginals of the biorthogonal KD distribution, which are expected to reflect the real probability distributions, straightforwardly provide the diagonal elements of \(\rho_{BO}\): \(\sum_{j=1}^{d} Q_{jk}^{BO} = \bar{\rho}_{kk}\) and \(\sum_{k=1}^{d} Q_{jk}^{BO} = \bar{\rho}_{jj}\).

\section{Genuine Quantum Superposition} \label{Sec::GS}

In Sec.~\ref{Sec::RTS::Mat}, we provide a comprehensive overview of the non-orthogonal matrix representation commonly employed for the density operator \(\hat{\rho}\) in the resource theory of superposition literature. We establish that this matrix, denoted as \(\rho_{N\!O}\), poses challenges in straightforward decomposition into probability and superposition components, as elaborated in Sec.~\ref{Sec::RTS::Phys}. Consequently, we delineate superpositions quantifiable through the off-diagonal elements of \(\rho_{N\!O}\), such as those measured by the \(l_1\) norm presented in Eq.~\eqref{Eq::l1norm::RTS}, as inter-basis superpositions. Furthermore, we discern superpositions resulting from basis state indistinguishability as intra-basis superpositions.

Moving to Sec.~\ref{Sec::BORTS}, we propose an alternative representation of the density operator \(\hat{\rho}\) using the framework of biorthogonality, denoted as \(\rho_{BO}\). We demonstrate that the elements of \(\rho_{BO}\) allow for decomposition into probabilities and superposition terms. Within this section, we conceptualize the interplay between intra- and inter-basis superpositions, leading to a comprehensive definition of superposition encompassing both. Building upon this conceptual framework, we explore potential extensions of existing superposition measures to accommodate this refined definition.

\subsection{General Concept} \label{Sec::GS::Intro}

We introduce the term \textit{genuine quantum superposition} to denote the overall nonclassicality of quantum state \(\hat{\rho}\) with respect to the nonorthogonal basis \(\{|c_i\rangle\}\), which includes the inter-basis and intra-basis superpositions as two subsets.

Before proceeding further, it is crucial to elucidate the interpretation of basis state indistinguishability within the context of intra-basis superposition. Each basis state \(|c_i\rangle\) inherently exhibits a superposition that precludes its perfect differentiation from other basis states. This is what results in the emergence of a residual superposition during the decomposition of state \(\hat{\rho}\) into the basis states using the POVM elements \(\{\hat{F}_i\}_{i=1}^{d+1}\) in Sec.~\ref{Sec::RTS::Phys}. To illustrate it, let us consider two nonorthogonal states \(|c_\mu\rangle\) and \(|c_\nu\rangle\). The indistinguishability of these states is associated with the magnitude of their overlap \(G_{\mu\nu} = \langle c_\mu|c_\nu \rangle\). When one of them is expressed as a superposition of the other state and its dual, the overlap manifests in both amplitudes as follows:
\begin{equation} \label{Eq::IntaBasisSup1}
  |c_\nu\rangle = G_{\mu\nu} |c_\mu\rangle + (1-|G_{\mu\nu}|^2) |c_\nu^\perp\rangle \,
\end{equation}
thus elucidating that basis state indistinguishability stems from the quantum superposition principle. The elements of \(\rho_{N\!O}\) do not contain \(G_{\mu \nu}\). However, information regarding this overlap is found among the elements of \(\rho_{BO}\) (see Eq.~\eqref{Eq::NO2BO}).

We can now explore why we incorporate intra-basis superposition into the concept of genuine superposition, alongside inter-basis superposition. Eq.~\eqref{Eq::IntaBasisSup1} can be rewritten as follows:
\begin{equation} \label{Eq::IntaBasisSup2}
  |c_\nu\rangle =  G_{\mu\nu} |c_\mu\rangle + (1-|G_{\mu\nu}|^2) \sum_{j=1}^d (G^{-1})_{j \nu} |c_j\rangle ,
\end{equation}
which implies that an inter-basis superposition of the form \(\psi_\mu |c_\mu \rangle + \psi_\nu | c_\nu \rangle \) is equivalent to \((\psi_\mu+\psi_\nu  G_{\mu\nu} )|c_\mu \rangle + \psi_\nu (1-|G_{\mu\nu}|^2) \sum_{j=1}^d (G^{-1})_{j \nu} |c_j\rangle\). Therefore, intra- and inter-superpositions can be transformed into each other.

In this particular context, let us reconsider the inter-basis superposition-free state provided in \eqref{Eq::SFreeState}:
\begin{equation} \label{Eq::SFreeState2}
\begin{aligned}
\hat{\rho}_f &= \sum_i p_i |c_i\rangle \langle c_i | \Big( \sum_j |c_j\rangle \langle c_j^\perp | \Big) \\
&= \sum_{ij} p_i G_{ij} |c_i\rangle \langle c_j^\perp | ,
\end{aligned}
\end{equation}
which indicates that it is not genuinely superposition-free. Consequently, its bipartition in \eqref{Eq::QCState} exhibits non-zero quantum discord, as illustrated in Fig.~\ref{Fig::GeoDiscord}. It should be noted that while \(\rho_{N\!O}\) representation of this state manifests as a diagonal matrix, the off-diagonal elements of \(\rho_{BO}\) are \(p_i G_{ij}\).

On the contrary, a genuinely superposition-free state represented by a diagonal \(\rho_{BO}\) is expected to adhere to the following expression:
\begin{equation} \label{Eq::GSFreeState}
\begin{aligned}
\hat{\rho}_{f_g} &= \sum_{i} p_i |c_i\rangle \langle c_i^\perp | = \sum_{ij} p_i (G^{-1})_{ij} |c_i\rangle \langle c_j | .
\end{aligned}
\end{equation}

Hence, a genuinely superposition-free state necessitates non-zero inter-basis superposition, as evidenced by the emergence of non-zero off-diagonal elements in \(\rho_{N\!O}\) representation of this state. This implies that genuine superposition cannot be decomposed additively into intra- and inter-superpositions. In other words, intra- and inter-superpositions possess the capacity to neutralize each other through interference.

\subsection{Quantification} \label{Sec::GS::Measures}

The probabilistic and non-probabilistic aspects of a density operator \(\hat{\rho}\) (see Sec.~\ref{Sec::Trace}) find representation in the diagonal and off-diagonal elements of its biorthogonal matrix \(\rho_{BO}\), as elaborated in Sec.~\ref{Sec::BORTS}. Thus, the genuine quantum superposition inherent in \(\hat{\rho}\) can be quantified by examining the off-diagonal elements of \(\rho_{BO}\), for instance, through the \(l_1\) measure defined below:
\begin{equation} \label{Eq::l1norm::GS}
l_1[\rho_{BO}] = \sum_{j\neq k} |\bar{\rho}_{jk}| .
\end{equation}

To quantify the intra-basis superposition accurately, we first need to reset all the inter-basis superpositions inherent in the density operator. Let's denote a mapping function by \(\Lambda\) that eliminates the off-diagonal elements from the input matrix. Consequently, the \(l_1\) norm defined below effectively captures the intra-basis superposition:
\begin{equation}
\label{Eq::l1norm::IntraS}
l_1[\Lambda(\rho_{N\!O})\, G] = \sum_{j\neq k} \Big|\sum_l  \Lambda(\rho_{N\!O})_{jl}\, G_{lk}\Big| .
\end{equation}

This equation measures the extent of intra-basis superposition within the state \(\hat{\rho}\), concerning the non-orthogonal basis \(\{|c_i\rangle\}\). To clarify, let us decompose the general state~(\ref{Eq::State::RTS}) as \(\hat{\rho} = \hat{\rho}_f + \hat{\chi}_\rho\) where \(\hat{\rho}_f\) is the inter-basis superposition-free state given in Eq.~(\ref{Eq::SFreeState}) and \(\hat{\chi}_\rho = \sum_{i\neq j} \rho_{ij} |c_i\rangle \langle c_j|\). Then, \(l_1[\Lambda(\rho_{N\!O})\, G]\) turns out to be \(\sum_{i\neq j} p_i |\langle c_i|c_j\rangle|\), which is nothing but the weighted sum of the overlaps between the basis states.

As elaborated in Sec.~\ref{Sec::GS::Intro}, while both intra-basis and inter-basis superpositions are constituents of genuine quantum superposition, their simple addition does not comprehensively depict the genuine superposition phenomenon. This suggests that \(l_1[\rho_{BO}]\) cannot always be additively dissected into \(l_1[\rho_{N\!O}]\) and \(l_1[\Lambda(\rho_{N\!O}])\, G]\). For the sake of simplicity and without loss of generality, let us initially consider a two-level system existing in the state
\begin{equation}
\begin{aligned}
\hat{\rho} = \mathcal{N} (p \, |c_1\rangle \langle c_1| &+ (1-p) |c_2\rangle \langle c_2| \\
&+ \lambda |c_1\rangle \langle c_2| + \lambda^* |c_2\rangle \langle c_1|) \, ,
\end{aligned}
\end{equation}
where \(\langle c_1 | c_2 \rangle = s\) and \(\mathcal{N} = 1/(1 + \lambda \, s^* + s \, \lambda^*)\). In this case, \(l_1[\rho_{N\!O}]\) and \(l_1[\Lambda(\rho_{N\!O})\, G]\) become \(2 \, \mathcal{N} |\lambda|\) and \(\mathcal{N}|s|\), respectively. On the other hand, \(l_1[\rho_{BO}]\) gives \(\mathcal{N}(|p \, s + \lambda|+|(1-p) s + \lambda|) \leq l_1[\rho_{N\!O}]\) + \(l_1[\Lambda(\rho_{N\!O})\, G]\), where the equality holds only for real positive \(s\) and \(\lambda\) values. It is straightforward to repeat the same calculations in higher dimensions as the off-diagonals of \(\rho_{BO}\) can be written as
\begin{equation}\label{Eq::l1norm::Genuine}
\bar{\rho}_{ij} = \rho_{ij} + \rho_{ii} G_{ij} + \sum_{k \neq i} \rho_{ik} G_{kj} \, ,
\end{equation}
where the first two terms are respectively the off-diagonals of \(\rho_{N\!O}\) and \(\Lambda(\rho_{N\!O})\, G\). The appearance of these two off-diagonals as a sum in \(\bar{\rho}_{ij}\) suggests an interference between the inter-basis and intra-basis superpositions. Contrarily, the last term inside Eq.~(\ref{Eq::l1norm::Genuine}) indicates a synergetic contribution to the genuine superposition. However, this term can also decrease or increase the \(l_1\) norm of genuine superposition according to its relative sign.

\section{Outlook}

In this paper, we present a comprehensive framework that unifies quantum superposition and basis state quantum indistinguishability, the latter being regarded as a specialized case of the former termed intra-basis superposition. Our approach integrates a resource-theoretical perspective, leveraging the concept of biorthogonality. To quantify intra-basis quantum superposition and genuine quantum superposition, we introduce the proper generalizations of the conventional \(l_1\) measure of quantum coherence. This enable us to claim that genuine quantum superposition is the fundamental notion of nonclassicality, which always includes quantum coherence and correlations as special subsets.

Our work can be naturally extended in the following directions. First, our approach may pave the way for new measures in the resource theory of superposition. In this regard, the pseudo-Hermiticity of biorthogonal matrix \(\rho_{BO}\) warrants further investigation. We indeed believe that such an investigation may provide alternative methods for the quantification of genuine quantum superposition.

Second, the measures that we introduced in this paper can be applied to quantum chemistry to quantify the nonclassicality in the phenomena of chemical bonding. We have recently initiated exploratory efforts in this direction~\cite{2023_Aromatiq}. In this parallel inquiry, we successfully quantified electron delocalization in aromatic molecules utilizing genuine quantum superposition. This was not possible using the previous framework of superposition theory, since the formation of a chemical bond between two atoms is closely associated with overlapping of nonorthogonal atomic orbitals.

Last, within the unified framework presented here, it might be possible to study the place of quantum superposition in the hierarchy of nonclassicality. In order to deal with the nonorthogonality of atomic orbitals, L\"{o}wdin developed a symmetric orthogonalization method in 1950~\cite{1950_Lowdin}. This method gives the closest orthogonal basis \(\{|l_i\rangle\}\) in the least-squares sense to the original nonorthogonal basis \(\{|c_i\rangle\}\) through the transformation
\begin{equation} \label{Eq::Transform2Lowdin}
|l_j \rangle = \sum_i (G^{-1/2})_{ij} |c_i\rangle ,
\end{equation}
where \(\{|l_i\rangle\}\) is called as the L\"{o}wdin basis. When the density operator is written in this orthogonal basis
\begin{equation} \label{Eq::State::Lowdin}
\hat{\rho} = \sum_{i,j} \tilde{\rho}_{ij} |l_i\rangle \langle l_j | ,
\end{equation}
the coefficients \(\tilde{\rho}_{ij} = \langle l_i | \hat{\rho} | l_j \rangle\) form a density matrix labeled by \(\rho_{LO}\). Let us call it the \textit{L\"{o}wdin matrix representation} of \(\hat{\rho}\).

It is straightforward to show that \(\rho_{LO} = G^{1/2} \rho_{N\!O} \, G^{1/2} = G^{1/2} \rho_{BO} \, G^{-1/2}\). As \(G_{ij}\) goes to \(\delta_{ij}\), \(G\) and its powers turn out to be \(\mathds{1}\). This means that \(\rho_{N\!O}\), \(\rho_{BO}\), and \(\rho_{LO}\) become identical in the limit of vanishing overlaps. Also, it is straightforward to show that the biorthonormal and L\"{o}wdin matrix representations share not only the same trace but also the same \(l_1\) norm in some limited cases. Another independent study recently uncovered that maximally coherent states transform into states exhibiting maximal inter-basis superposition through the application of the L\"{o}wdin transformation~\cite{2023_GT_LO}. Can one generalize the L\"{o}wdin transformation to derive an operation that converts all the genuine quantum superposition into quantum coherence? We believe this may lead us toward a complete theory of nonclassicality which puts superposition, coherence, and discord on a unified standing.

\begin{acknowledgments}	

I am grateful to Ali Y{\i}ld{\i}z, Vlatko Vedral, and Martin Plenio for providing a tremendous amount of encouragement and suggestions. I would like to thank Ali Mostafazadeh and Nicole Yunger Halpern for having introduced me to the literatures of pseudo-Hermiticity and Kirkwood-Dirac distribution. I would also like to acknowledge my particular indebtedness to G\"{o}khan Torun for useful suggestions and extensive discussions. This work was supported by the Scientific and Technological Research Council of Turkey (T\"{U}B\.{I}TAK) under Grant No. (120F089).

\end{acknowledgments}

\end{document}